\documentclass[numberedappendix]{emulateapj}
\usepackage{apjfonts,mathptmx}
\usepackage{pstricks,pst-grad,multido}

\begin{document}

\title{Spectral Methods for Time-Dependent Studies of Accretion Flows. III.\\
       Three-Dimensional, Self-gravitating, Magnetohydrodynamic Disks}

\author{Chi-kwan Chan, Dimitrios Psaltis\altaffilmark{1}, and Feryal \"Ozel}

\affil{Physics Departments, University of Arizona, 1118 E.\ 4th St.,
       Tucson, AZ 85721}

\altaffiltext{1}{Also Astronomy Department, University of Arizona}

\begin{abstract}
  Accretion disks are three-dimensional, turbulent, often
  self-gravitating, magnetohydrodynamic flows, which can be modeled in
  detail with numerical simulations.  In this paper, we present a new
  algorithm that is based on a spectral decomposition method to
  simulate such flows.  Because of the high order of the method, we
  can solve the induction equation in terms of the magnetic potential
  and, therefore, ensure trivially that the magnetic fields in the
  numerical solution are divergence free.  The spectral method also
  suffers minimally from numerical dissipation and allows for an easy
  implementation of models for sub-grid physics.  Both properties make
  our method ideal for studying MHD turbulent flows such as those
  found in accretion disks around compact objects. We verify our
  algorithm with a series of standard tests and use it to show the
  development of MHD turbulnce in a simulation of an accretion disk.
  Finally, we study the evolution and saturation of the power spectrum
  of MHD turbulence driven by the magnetorotational instability.
\end{abstract}



\section{Introduction}

Although the standard accretion disk model was proposed more than
thirty years ago \citep{Shakura1973}, the properties of turbulent
angular momentum transport in accretion disks are still not well
understood.  Shakura \& Sunyaev (1973) hypothesized in their original
work that magnetic fields may be important in mediating the required
angular momentum transport.  However, it was not until the last decade
that~\citet{Balbus1991a, Balbus1991b} pointed out that the
magnetorotational instability (MRI) generates turbulence and leads to
transport of angular momentum in accretion disks.

The non-linear evolution of the MRI and the generation of turbulence
was studied numerically by~\citet{Balbus1991b}, following the earlier
linear analysis of the instability.  Later, local numerical
simulations were performed in the shearing box approximation
\citep[e.g.,][]{Hawley1995, Brand1995, Hawley1996}, aimed to study
further the local properties of three-dimensional MRI, with and
without stratification~\citep{Stone1996}.  The natural extension of
shearing box calculations, namely the cylindrical disks with vanishing
vertical gravitational force, were simulated by~\citet{Hawley2001a,
Armitage2001} to illustrate some important aspects of the turbulent
transport, especially in the vicinity of the innermost stable circular
orbit around a black hole.  Finally, global numerical simulations of
MHD disks have also been carried out for a variety of settings and
physical conditions.

All of the codes that have been used to study the properties of
MRI-driven turbulence have been based on two types of differencing
schemes.  The first class of studies make use of the very successful
scheme developed originally for the ZEUS code (e.g., Stone \& Pringle
2001; Armitage et al. 2001; see Stone \& Norman 1992a and 1992b for
the ZEUS code) or schemes based on it (e.g., Hawley 2000; De Villiers
\& Hawley 2003; Steinacker \& Papaloizou 2002; Igumenschev et
al. 2003).  The other class of studies have used different
conservative schemes (e.g., Koide et al. 1999; Gammie et al. 2003;
Machida \& Matsumoto 2003).  Both types of finite difference methods
allow for a stable and efficient implementation of solvers of the MHD
equations.  However, they also introduce a considerable amount of
numerical dissipation to the problem.  This is significant because
most calculations have been performed for ideal MHD and, hence, it is
this numerical dissipation that allows for the MRI instability to
saturate and the resulting turbulence to reach a dynamical steady
state (see, however, Flemming, Stone, \& Hawley 2000 where the
assumption of ideal MHD is relaxed).  As a result, the kinetic and
magnetic energies of different simulations saturate at different
levels depending on the resolution and the scheme \citep{Hawley1999}.
The effect of this shortcoming can be reduced, e.g., by increasing the
resolution, by increasing the discretization order, or by using
numerical schemes that reduce numerical diffusion \citep[see,
e.g.,][for an unsplit Godunov MHD code]{Gardiner2005}.

In this third paper of the series, we address this issue by developing
a version of our pseudo-spectral numerical algorithm to simulate
three-dimensional MHD disks.  Spectral algorithms are high order
numerical methods, in which dynamical variables are evolved along
orthogonal modes.  For smooth functions, they require only $\sim\pi$
grid points to accurately resolve one wavelength, compared to $\sim
16$ grid points for finite difference schemes to reach the same
accuracy.  Moreover, the MHD equations can be evolved in time in terms
of the vector potential $\mathbf A$, preserving thus trivially the
cylindrical character of the magnetic field.

The high order of spectral methods makes them ideal for studying
problems of magnetohydrodynamic turbulence, since they do not suffer
from serious numerical dissipation.  Moreover, spectral methods can
easily incorporate models of sub-grid physics, such as the those
involved in large-eddy simulation~\citep[see, e.g.,][]{Blackburn2003}.
The idea of large-eddy-simulations is to model approximately the
small-scale structures, instead of resolving all features of a
turbulent flow~\citep[see, e.g.,][]{Sagaut2004}, with sub-grid models
that are based either on experiments or phenomenological descriptions
of the small-scale physics.  The large-eddy-simulation approach fits
naturally to spectral methods because the scale-separation operation
is mathematically a frequency low-pass filter, which is easily
incorporated in a spectral scheme.


In the next section we present the system of equations we solve.  In
particular, we describe our treatment of the induction equation in
terms of the vector potential, in which we introduce an appropriate
gauge to avoid the unbounded (linear) growth in the vector potential
in a Keplerian flow.  In \S\ref{sec:methods} we describe the numerical
technique we use.  In \S\ref{sec:tests}, we present a series of tests
to verify the implementation of our algorithm and conclude, in
\S\ref{sec:disks}, with an application of our method to the study of
MHD turbulence in accretion disks driven by the magnetorotational
instability.


\section{Equations and Assumptions}
\label{sec:equations}

\emph{Magnetohyrodynamics.---\/}We consider three dimensional viscous,
compressible, magnetohydrodynamic flows.  The MHD equations contain
four equations, namely, the continuity equation
\begin{equation}
  \frac{\partial\rho}{\partial t} + \nabla\cdot(\rho\mathbf v) = 0,
  \label{eq:continuity}
\end{equation}
the momentum equation
\begin{equation}
  \rho\frac{\partial\mathbf v}{\partial t} + \rho(\mathbf
  v\cdot\nabla) \mathbf v = -\nabla\left(P + \frac{B^2}{8\pi}\right) +
  \frac{1}{4\pi}(\mathbf B \cdot \nabla)\mathbf B + \nabla\mathbf\tau
  + \rho\,\mathbf g,
  \label{eq:navier_stokes}
\end{equation}
the energy equation
\begin{equation}
  \frac{\partial E}{\partial t} + \nabla\cdot(E\mathbf v) = -
  P\nabla\cdot\mathbf v + \Phi_\nu + \Phi_B - \nabla\cdot\mathbf q -
  \nabla\cdot\mathbf F,
  \label{eq:energy}
\end{equation}
and the induction equation
\begin{equation}
  \frac{\partial\mathbf A}{\partial t} = \mathbf v \times
  (\nabla\times\mathbf A) + \frac{c^2}{4\pi\sigma}\nabla^2\mathbf A 
  + \nabla\dot\Lambda.
  \label{eq:induction}
\end{equation}
We denote by $\rho$ the density, by $\mathbf v$ the velocity, and by
$E$ the thermal energy.  In the momentum equation, $P$ is the thermal
pressure, $\mathbf\tau$ is the viscosity tensor, and $\mathbf g$ is
the gravitational acceleration.  In the energy equation, there are two
dissipative terms, namely, the Ohmic dissipation $\Phi_\eta$ and the
viscous dissipation $\Phi_\nu$.  We use $\mathbf q$ to denote the heat
flux vector and $\mathbf F$ to denote the radiation flux.

We write the induction equation in terms of the vector potential
$\mathbf A$, so that the magnetic field is given by $\mathbf B =
\nabla\times\mathbf A$.  The symbol $\sigma$ here represents the
electrical conductivity and we define the microscopic resistance by
$\eta \equiv c^2/4\pi\sigma$.  The last term, $\nabla\dot\Lambda$, in
the induction equation is a gauge source/sink term, the purpose of
which we explain below.

It is straightforward to show that equation~(\ref{eq:induction}) leads
to the standard induction equation
\begin{equation}
  \frac{\partial\mathbf B}{\partial t} = \nabla\times(\mathbf v \times
  \mathbf B) + (\nabla\cdot\eta\nabla)\mathbf B\;.
\end{equation}
Because $\nabla\times(\nabla\dot\Lambda) \equiv 0$,
$\nabla\dot\Lambda$ does not affect the magnetic field.  However, we
retain this gauge term because, by proper choice, it can be used to
suspend a non-physical (i.e., numerical) linear growth in $\mathbf A$
and hence improve the accuracy of the scheme.  To illustrate this, we
consider a Keplerian disk with a constant vertical magnetic field so
that the vector $\mathbf v\times\mathbf B$ has a non-zero
$r$-component.  By assuming $\eta = 0$, the potential form of the
induction equation reduces to
\begin{equation}
  \frac{\partial A_r}{\partial t} = B_z \sqrt{\frac{GM}{r}} +
  \frac{\partial\dot\Lambda}{\partial r},
\end{equation}
where $G$ and $M$ are the gravitational constant and the mass of the
central object, respectively.  The first term on the right hand side
leads to a linear growth of $ A_r$ in time.  This growth will never
saturate, because the mean of the product $v_\phi B_z$ is always
positive.  Although the actual value of the magnetic field will not be
affected, this growth will lead to a large round off error in $A_r$ if
the MHD equations are integrated for a long time.  This difficulty can
be overcome by setting
\begin{equation}
  \dot\Lambda \equiv -B_z \int dr \sqrt{\frac{GM}{r}}
  = -2 B_z\sqrt{GMr},
\end{equation}
which suppresses the linear growth of the vector potential.  In our
algorithm, $\dot\Lambda$ is calculated dynamically from the values of
$\bar B_z(t,r)$ and $\bar v_\phi(t,r)$, where the over-bars indicate
averages over the azimuthal and vertical directions of the disk,
respectively.

The analytical forms of the various physical quantities in
equations~(\ref{eq:continuity}) --- (\ref{eq:energy}) were given in
\citet{Chan2005}. Here, we generalize them to three dimensions.  The
viscosity tensor (in Cartesian coordinates) is
\begin{equation}
  \tau_{ij} = 2(\mu_\mathrm{r} + \mu_\mathrm{s})e_{ij} +
  \left(\mu_\mathrm{r} + \mu_\mathrm{b} -
  \frac{2}{3}\mu_\mathrm{s}\right)(\nabla\cdot\mathbf v)\delta_{ij},
\end{equation}
where the strain-rate tensor $e_{ij}$ is
\begin{equation}
  e_{ij} = \frac{1}{2}\left(\frac{\partial v_i}{\partial x_j} +
  \frac{\partial v_j}{\partial x_i}\right).
\end{equation}
The viscous dissipation rate is, therefore,
\begin{equation}
  \Phi_\nu = 2(\mu_\mathrm{r} + \mu_\mathrm{s})(e_{ij})^2 +
  \left(\mu_\mathrm{r} + \mu_\mathrm{b} -
  \frac{2}{3}\mu_\mathrm{s}\right)(\nabla\cdot\mathbf v)^2.
\end{equation}
Finally, the Ohmic dissipation rate, $\Phi_\eta$, is given by
\begin{equation}
  \Phi_\eta = \frac{J^2}{\sigma} = \frac{\eta}{4\pi} |\nabla\times\mathbf B|^2.
\end{equation}
We again assume an ideal gas law so that
\begin{eqnarray}
  E & = & \rho\frac{3k_\mathrm{B}T}{2\mu m_\mathrm{H}}, \\
  P & = & \rho\frac{k_\mathrm{B}T}{\mu m_\mathrm{H}}.
\end{eqnarray}
For the induction equation, we typically set the resistance $\eta$ to
zero so that the diffusion term in the induction equation vanishes.

\emph{Gravity.---\/}We solve for the gravitational acceleration,
$\mathbf g$, in a similar way as in \citet{Chan2006}.  We first define
the gravitational potential $\Psi$ by
\begin{equation}
  \mathbf{g} \equiv -\nabla\Psi,
\end{equation}
which is given by the volume integral
\begin{equation}
  \Psi(t,\mathbf{x}) = -G\int\frac{\rho(t,\mathbf{x'})}{|\mathbf{x} -
  \mathbf{x'}|}d^3x'
  \label{eq:integrator_3d_total}
\end{equation}
over all space.  Rewriting equation~(\ref{eq:integrator_3d_total}) in
differential form, we obtain Poisson's equation
\begin{equation}
  \nabla^2\Psi = 4\pi G\rho,
\end{equation}
with $\Psi$ satisfying the boundary condition $\Psi(t,\infty) = 0$ at
all times.  When simulating accretion flows, the computational domain
$\mathcal{D}^{(3)}$ is usually finite.  Based on its linearity, we can
decompose Poisson's equation into two parts, i.e.,
\begin{equation}
  \nabla^2\Psi_\mathrm{int} = 4\pi G\rho_\mathrm{int},
  \label{eq:possion_int} 
\end{equation}
and
\begin{equation}
  \nabla^2\Psi_\mathrm{ext} = 4\pi G\rho_\mathrm{ext},
\end{equation}
where $\rho_\mathrm{int}$ denotes the mass density within the
computational domain, which in our case is the disk density, and
$\rho_\mathrm{ext}$ refers to external sources such as the central
object and/or a companion star.  The gravitational field is then given
by
\begin{equation}
  \mathbf{g} = \mathbf{g}_\mathrm{int} + \mathbf{g}_\mathrm{ext} =
  -\nabla(\Psi_\mathrm{int} + \Psi_\mathrm{ext}).
  \label{eq:total_g}
\end{equation}
For the gravitational field of the central object, we use the
pseudo-Newtonian approximation of \citet{Mukhopadhyay2002} for
$\mathbf{g}_\mathrm{ext}$, which takes the form
\begin{equation}
  \mathbf g_\mathrm{ext} = - \frac{c^2}{r^3} \left(\frac{GM}{c^2}\right)^2
  \left[\frac{r^2 - 2(a/c)\sqrt{GMr/c^2} + (a/c)^2}{\sqrt{GMr/c^2}(r -
  r_\mathrm{S}) + a/c}\right]^2 \mathbf{\hat r}.
\end{equation}
Here, $r_\mathrm{S} \equiv 2 GM/c^2$ is the Schwarzschild radius and
$a$ is a parameter related to the spin of the central object.

In order to solve for self-gravity using
equation~(\ref{eq:possion_int}) within $\mathcal{D}^{(3)}$, we compute
the integral
\begin{equation}
  \Psi_\mathrm{int}(t,\mathbf{x}) = -G\int_{\mathcal{D}^{(3)}}
  \frac{\rho_\mathrm{int}(t,\mathbf{x'})}{|\mathbf{x} -
  \mathbf{x'}|}d^3x'\;,
  \label{eq:integrator_3d}
\end{equation}
as in~\citet{Chan2006}.  We then use equation~(\ref{eq:total_g}) to
obtain the total gravitational field and use it in the momentum
equation.

\emph{Subgrid Physics.---\/}For homogeneous and isotropic turbulence,
the ratio between the most energetic scale to the viscous length scale
is proportional to $\mathcal{O}(Re^{3/4})$, where $Re$ is the Reynolds
number.  This scaling law suggests that the computational cost is
proportional to $\mathcal{O}(Re^3)$ for three-dimensional,
time-dependent simulations.  Although spectral methods offer efficient
ways to capture small scale features, they are still unable to
resolve, even in the shearing box approximation, the expected
dynamical range in accretion disks down to the molecular viscous
length scale.

In order to reduce the computational cost, we need to introduce an
artificial cut-off to the problem. Following the Large Eddy Simulation
approach~\citep[LES, see][for a very detailed
introduction]{Sagaut2004}, we will use our numerical algorithm to
solve the MHD equations for scales larger than this cut-off scale and
introduce an approximate model for the smaller scales.  Introducing
such a model is required by the non-linear character of the MHD
equations, which allow for coupling between the simulated large scales
and the unresolved small scales.

For any physical quantity $f$, we denote by $\overline{f}$ the
filtered (i.e., resolved) function and define the sub-grid fluctuation
by
\begin{equation}
  f' \equiv f - \overline{f}.
\end{equation}
We can then formally decompose any non-linear product of two physical
quantities $fg$ in physical space as
\begin{eqnarray}
  \overline{fg} & = & \overline{(\overline{f}+f')(\overline{g}+g')} \nonumber\\
  & = & \overline{f}\overline{g}+L_{ij}+C_{ij}+R_{ij}\;,
\end{eqnarray}
where
\begin{eqnarray}
  L_{ij} & = & \overline{\overline{f}\overline{g}} -
    \overline{f}\overline{g},\nonumber\\
  C_{ij} & = & \overline{\overline{f} g'}+\overline{f'\overline{g}},
    \nonumber \\
  R_{ij} & = & \overline{f' g'}.
\label{eq:SEL}
\end{eqnarray}
If the quantities $f$ and $g$ are different components of the
velocity, these higher order correlations are the Leonard tensor, the
cross-stress tensor, and the Reynolds sub-grid tensor, respectively
\citep{Sagaut2004}.  Note that the Reynolds sub-grid tensor is
different from the (physical) Reynolds stress tensor.  It appears
because of the presence of the artificial cut-off and is independent
of the molecular viscosity.

The basic idea of Large Eddy Simulations is to devise a model for the
three tensors~(\ref{eq:SEL}) that captures the physics of sub-grid
turbulence.  This is beyond the scope of the current paper. Here, we
will approximate the sum $L_{ij} + C_{ij} + R_{ij}$ in spectral space
by the the spectral filter (Chan et al.\ 2005)
\begin{equation}
  \sigma_\beta\left(\frac{n}{N}\right) =
  \exp\left( - |\ln\epsilon| \left|\frac{n}{N}\right|^\beta \right)\;,
\end{equation}
where $\beta$ is the order of the filter; $\epsilon$ is the machine
accuracy, which is of order $10^{-15}$ for double precision floating
point numbers; and $n$ and $N$ are the point index and number of
points in spectral space, respectively.  Applying this filter once
after every timestep is equivalent to adding a high-order (super)
diffusion term in the dynamic equations. Although this approach is not
based on any physical model, it has been known to reproduce the
large-scale properties of turbulent
flows~\citep[e.g.,][]{Karamanos2000, Pasquetti2005}.


\section{Implementation of the Pseudo-Spectral Methods}
\label{sec:methods}

In our current implementation of the pseudo-spectral method in
three-dimensions we use cylindrical coordinates because we are
interested in the study of geometrically thin accretion disks. It is
trivial, however, to alter the geometry of the domain of solution,
when necessary, and use spherical-polar coordinates.

Along the radial direction, we use a Chebyshev collocation method,
whereas, for the azimuthal and vertical directions, we choose the
Fourier basis.  This implies periodic boundary conditions for both the
azimuthal direction (which is natural) and the vertical direction
(which needs to be justified for each specific application).  The
computational domain is $\mathcal{D}^{(3)} =
[r_{\min},r_{\max}]\times[-\pi,\pi)\times[-Z, Z)$ and we require
$r_{\min} > 0$ to avoid the coordinate singularity at the origin.

Formally, we expand any physical quantity $f(t,r,\phi,z)$ as
\begin{equation}
  f(t,r,\phi,z) = \sum_{n,m,l} \check f_{nml}(t) T_n(\bar r)
  e^{im\phi} e^{i\pi lz/Z}.
\end{equation}
Here $T_n$ is the $n$-th order Chebyshev polynomial and $\bar r \in
[-1,1]$ is the standardized coordinate in the radial direction
\citep{Chan2005}.  Note that the frequently used
Chebyshev-Gauss-Lobatto grid
\begin{equation}
  \bar r_k \equiv \cos\left(\frac{\pi k}{N}\right)
  \mbox{, for $0 \le k \le N$}
\end{equation}
has the property that
\begin{equation}
  \bar{r}_0 - \bar{r}_1 = \bar{r}_{N-1} - \bar{r}_N \propto N^{-2}.
\end{equation}
This gives the time-stepping constrain $\Delta t \propto N^{-2}$ for
hyperbolic (wave-like) equations and $\Delta t \propto N^{-4}$ for
parabolic (diffusion-like) equations.  To overcome this restriction,
we use instead the Kosloff-Tal-Ezer mapping \citep{Kosloff1993}
\begin{equation}
  r = \frac{r_{\max}}{2}\left[\frac{\arcsin(\alpha\bar r)}{\arcsin(\alpha)}
  + 1\right]
  - \frac{r_{\min}}{2}\left[\frac{\arcsin(\alpha\bar r)}{\arcsin(\alpha)}
  - 1\right].
\end{equation}
We choose the parameter $\alpha$, which controls the regularity of the
grid spacing, to be $\mathrm{sech}(|\ln\epsilon|/N)$, where $\epsilon$
is the machine accuracy, to optimize the accuracy of the spatial
derivatives~\citep{Don1997}.

We compute the numerical derivatives along the $r$- and
$\phi$-directions as in \citet{Chan2005,Chan2006}.  Hence, the radial
derivative is given by the chain rule
\begin{equation}
  \frac{\partial f}{\partial r} = \frac{1}{dr/d\bar r}\frac{\partial
    f}{\partial\bar r},
\end{equation}
where the derivative in the standardized coordinate is
\begin{equation}
  \frac{\partial f}{\partial\bar r} = \sum_{n,m,l} \check{f}^{(1)}_{nml}(t)
  T_n(\bar r) e^{im\phi} e^{i\pi l z/Z}.
\end{equation}
We precompute analytically the derivative of the mapping, $dr/d\bar
r$.  Here, we use $\check f^{(1)}$ to denote the Chebyshev coefficient
of the radial derivative and compute it using the following three-term
recursive relation
\begin{eqnarray}
  \check{f}^{(1)}_{N,m,l} & = & 0, \\
  \check{f}^{(1)}_{N-1,m,l} & = & 2 N \check{f}_{N,m,l}, \\
  c_n \check{f}^{(1)}_{n,m,l} & = & \check{f}^{(1)}_{n+2,m,l}
  + 2(n+1) \check{f}_{n+1,m,l},
\end{eqnarray}
where $c_0 = 2$ and $c_n = 1$ for $n=1, 2, \dots, N$.  The azimuthal
derivative is given by
\begin{equation}
  \frac{\partial f}{\partial\phi} = \sum_{n,m,l}
  im\check{f}_{nml}(t) T_n(\bar r) e^{im\phi} e^{i\pi l z/Z}.
\end{equation}
Finally, the derivative along the $z$-direction is similar to the
azimuthal derivative except for the extra normalization $\pi/Z$, i.e.,
\begin{equation}
  \frac{\partial f}{\partial z} = \sum_{n,m,l} \frac{i\pi
  l}{Z} \check f_{nml}(t) T_n(\bar r) e^{im\phi} e^{i\pi lz/Z}.
\end{equation}


\section{Code Verification}
\label{sec:tests}

We have verified our numerical algorithm using a suit of test
problems, some of which we present in this section.  For a test
particular to the three-dimensional hydrodynamics, we adopt the
free-falling dust ring test from \citet{Chan2005}; for MHD, we study
the magnetic braking of a rotating slab, following \citet{Stone1992}.


\subsection{An Advection Test: Free Fall of a Dust Ring}
\label{sec:free_fall}

\begin{figure}
  \plotone{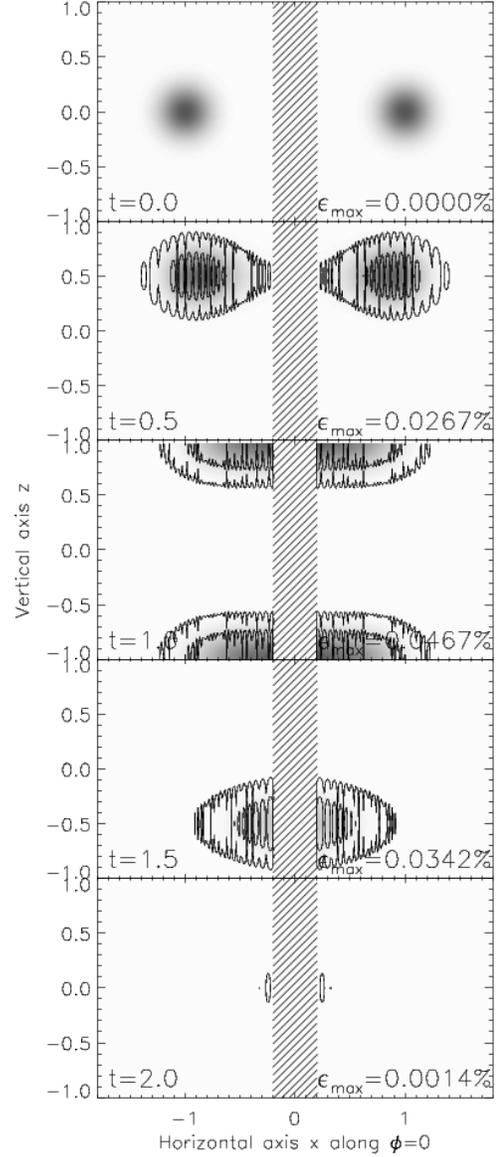}\\
  \caption{A comparison of the numerical to the analytical solution
    for the free-falling dust ring test problem discussed in the text.
    The gray-scale image depicts the density profile at different
    times in the simulation.  The contour lines show the positions in
    the solution domain where the fractional error between the
    numerical and analytical solutions are $0.01\%$ and $0.001\%$. In
    all plots, we also show the maximum error $\epsilon_{\max}$ for
    reference.}  \label{fig:free_fall} \mbox{}
\end{figure}

Following \citet{Chan2005}, we use a free falling dust ring as an
advection test of our three-dimensional algorithm.  The computational
domain is $[0.2,1.8]\times[-\pi,\pi]\times[-1,1]$ with
$65\times32\times65$ collocation points. The initial density is the
Gaussian
\begin{equation}
  \rho_0 = \exp[-20(r-1)^2-20z^2].
\end{equation}
In order to also test the advection along the $z$-direction, we set
the initial velocity to $\mathbf v = (0, 0, 1)$, which is equivalent
with performing this test on a non-stationary Galilean frame.  We
assume that the gravitational acceleration is Newtonian and that of a
central external object, i.e., that $\mathbf g = (-1/r^2, 0,
0)$. (Note that, for a cylindrically symmetric central object, the
gravitational field should have been proportional to $1/r$.) We also
neglect pressure and magnetic fields in this test.

The analytical solution of this test is found in the same way as is
\citet{Chan2005}.  Because the initial vertical velocity is non-zero,
we have to replace $z$ by $z - v_z t$ in order to capture the vertical
motion.  The analytical solution, therefore, reads
\begin{equation}
  \Sigma(t,r,\phi,z) = \Sigma_0(r_0, z - v_z t) \frac{r_0^2}{r}
  \left[\frac{3t}{2}\sqrt{\left(\frac{2}{r_0}\right)
  \left(\frac{r_0}{r}-1\right) + r}\right]^{-1}
\end{equation}
where $r_0$ is found by solving implicitly the equation
\begin{equation}
  \frac{\sqrt{2}}{r_0^{3/2}}\;t =
  \frac{1}{2}\sin\left(2\arccos\sqrt{\frac{r}{r_0}}\right) +
  \arccos\sqrt{\frac{r}{r_0}}\;.
\end{equation}

\begin{figure*}
  \plottwo{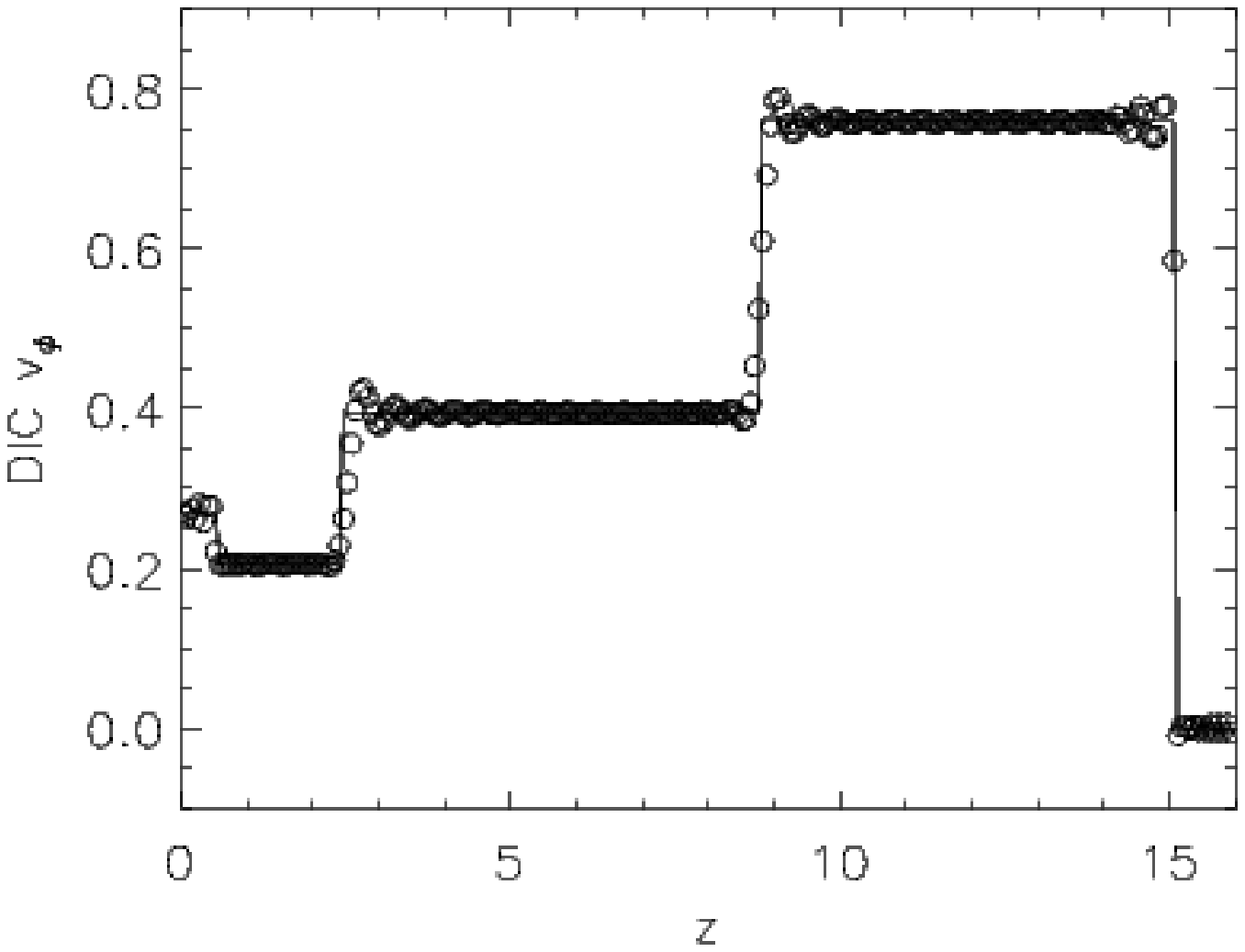}{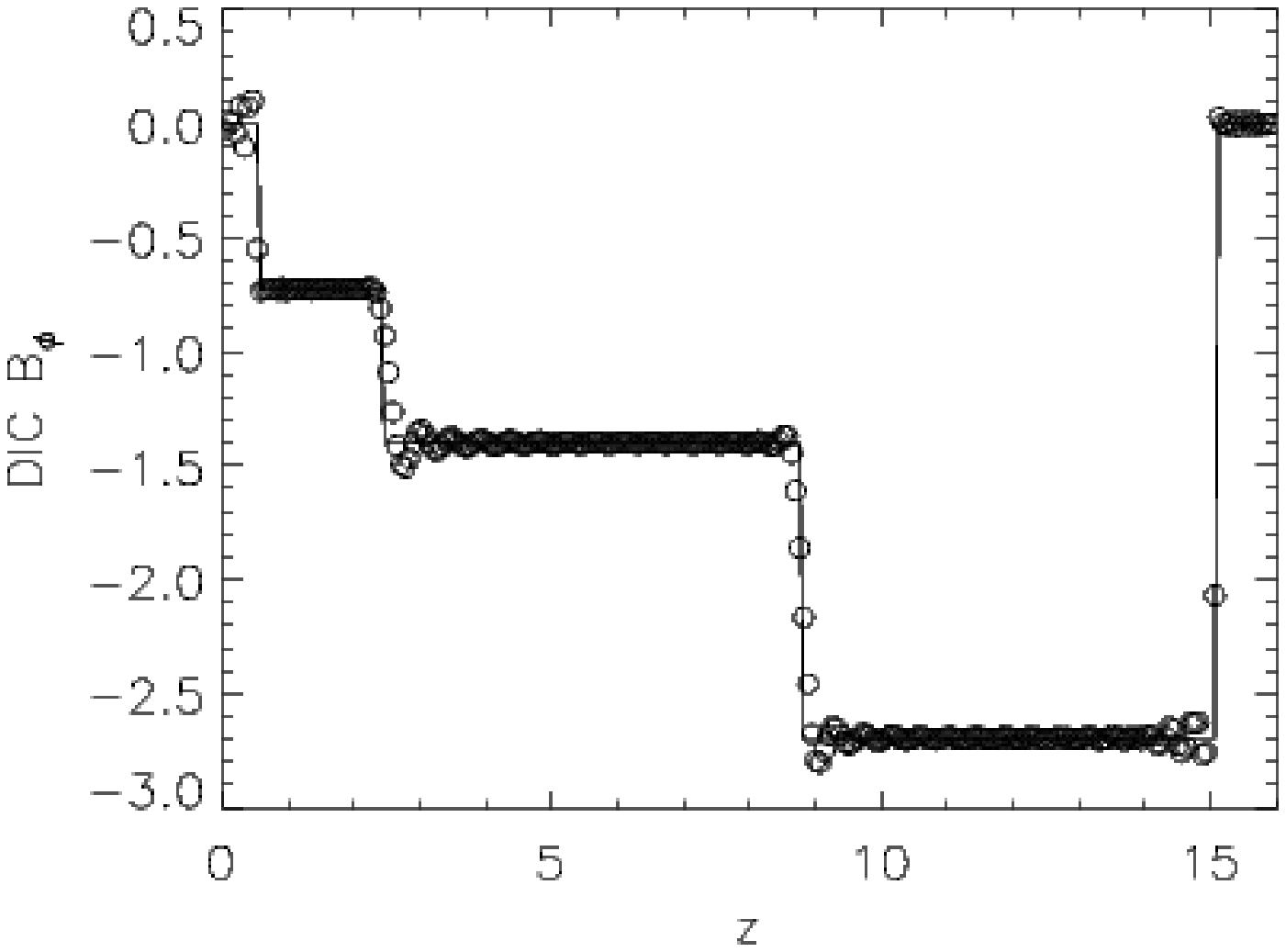}\\
  \plottwo{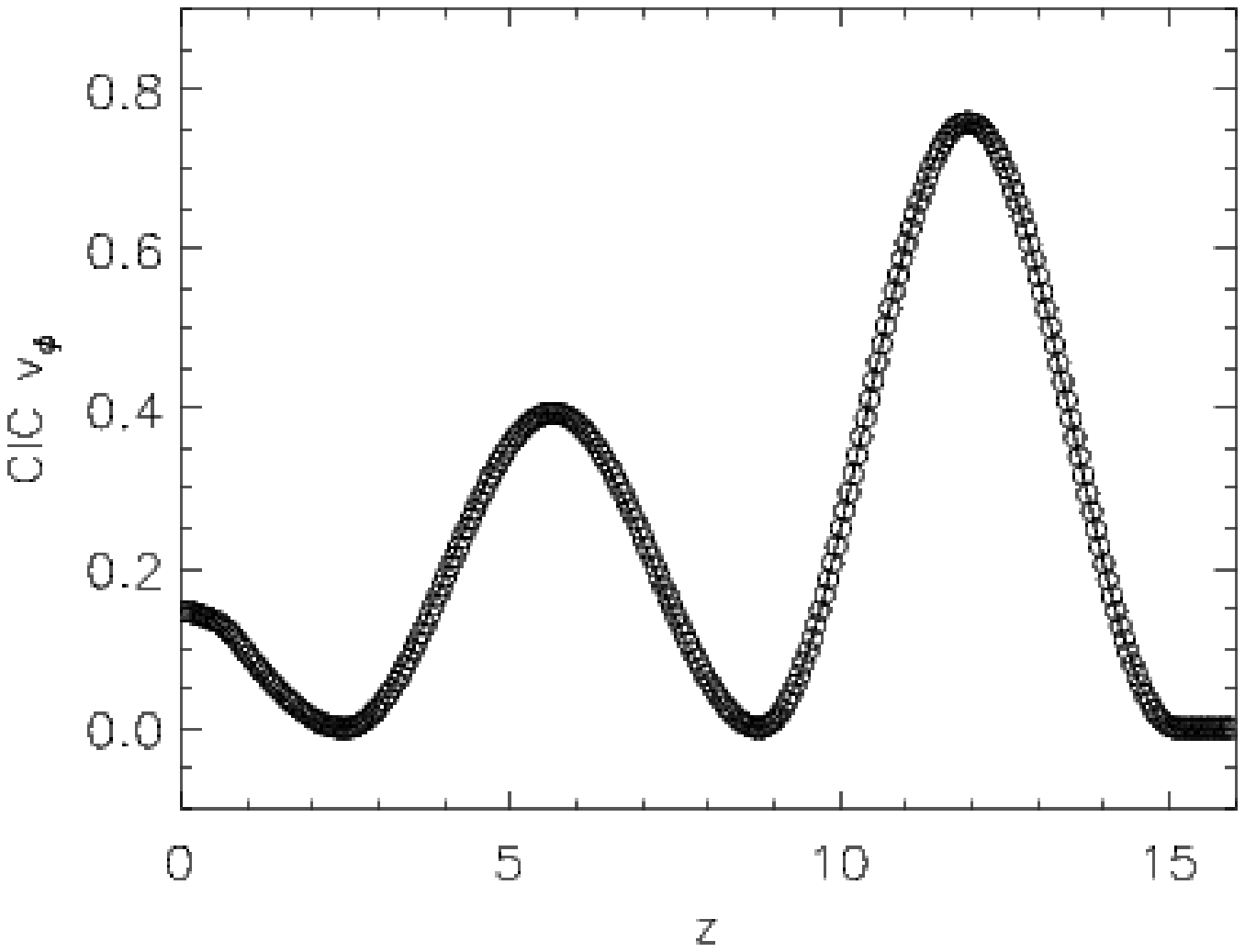}{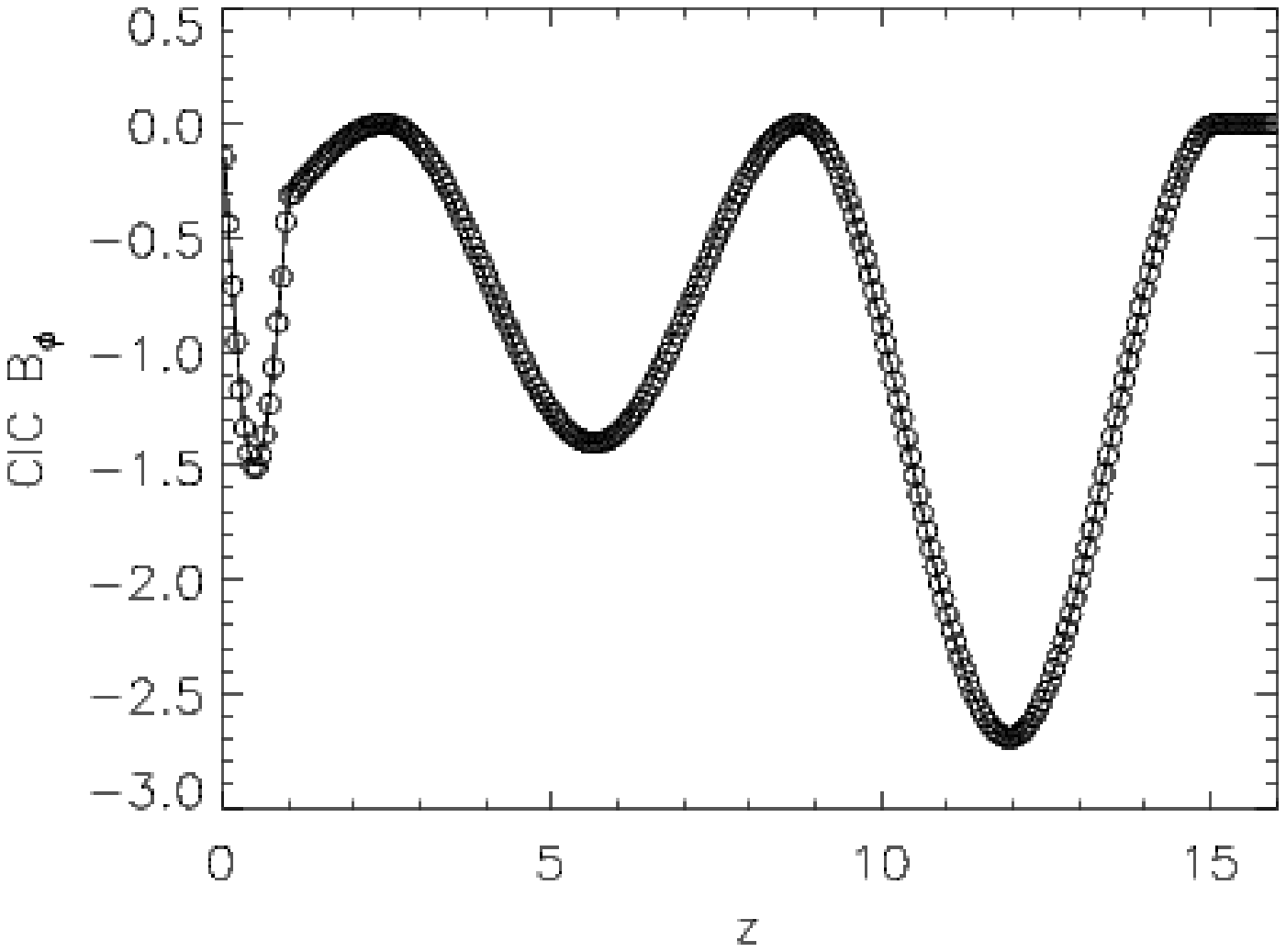}\\
  \caption{A comparison of the numerical (circles) to the analytical
    (lines) solutions for the magnetic breaking test problem. The
    panels show a snapshot of the azimuthal velocity and magnetic
    field at time $t=50$, for the discontinuous (DIC; upper panels)
    and the continuous (CIC; lower panel) initial conditions discussed
    in the text. For the discontinuous case, the numerical solution
    captures correctly the shock properties despite the Gibbs
    oscillations; for the continuous case, the numerical and
    analytical solutions are indistinguishable.}
\label{fig:braking}
\end{figure*}

In Figure~\ref{fig:free_fall}, we plot the numerical density as
gray-scale contours at $\phi = 0$ and for different times.  The error
between the numerical solution and the analytical solution is
overplotted as a set of contour lines. The maximum error throughout
the simulation is of order $10^{-4}$. Note that our implementation of
the free-streaming inner boundary condition does not introduce any
significant errors caused by the artificial excitation or reflection
of waves.

\mbox{}


\subsection{A Test of Alfven Wave Propagation: Magnetic Braking of an
Aligned Rotator}

The problem of magnetic breaking of an aligned rotator via the
emission of nonlinear, incompressible Alfv\'en waves was solved
analytically by \citet{Mouschovias1980}.  It was then used by
\citet{Stone1992} to verify the ability of MHD algorithms to propagate
transverse Alfv\'en waves. In this test, we use the same initial
conditions as those described by \citet{Stone1992}, namely, the
discontinuous initial condition (DIC) and the continuous initial
condition (CIC).  Because our algorithm is designed for compressible
MHD flows, we discard the radial and vertical components of the
momentum equation and set $\partial P/\partial\phi = 0$ in the
azimuthal component.

We use the computational domain $[0.2,1.8] \times [-\pi,\pi) \times
[-16,16]$ with $33 \times 32 \times 512$ collocation points.  We set
the initial density to
\begin{equation}
  \rho_0 = \left\{\begin{array}{lll}
    1  & , & |z| > 1 \\
    10 & , & \mbox{otherwise}
  \end{array}\right.
\end{equation}
the initial magnetic field to ${\mathbf B} = (0, 0, 1)$, and the
initial velocity to $\mathbf v = (0, r\Omega_0, 0)$ where $\Omega_0$
is the initial angular velocity. For the discontinuous case (DIC), we
set the angular velocity to
\begin{equation}
  \Omega_0 = \left\{\begin{array}{lll}
    0 & , & |z| > 1 \\
    1 & , & \mbox{otherwise,}
  \end{array}\right.
\end{equation}
whereas for the continuous case (CIC) we set it to
\begin{equation}
  \Omega_0 = \left\{\begin{array}{lll}
    0                        & , & |z| > 1 \\
    \frac{1}{2}(1+\cos\pi z) & , & \mbox{otherwise.}
  \end{array}\right.
\end{equation}
The analytic solutions are given in \citet{Mouschovias1980}.

In Figure~\ref{fig:braking}, we compare the numerical to the
analytical solutions for the two initial conditions at $t = 50$, right
before the wave front passes $z = 16$.  The numerical solution for the
discontinuous problem shows oscillations around the discontinuity (the
Gibbs phenomenon), which are inherent to all spectral methods.
However, the properties of the shock as well as those of the fluid
around it are captured correctly in our numerical solutions. For the
continuous problem, the numerical and analytical solutions are
indistinguishable.


\section{Turbulent MHD Disks Driven by the Magneto-Rotational Instability}
\label{sec:disks}

One of the advantages of spectral methods is the fact that they allow
for an accurate control of numerical dissipation and hence they can be
used to track accurately the stability of a flow.  In \citet{Chan2005}
we used our two-dimensional, hydrodynamic spectral algorithm to
successfully reproduce the Rayleigh stability criterion in a couette
flow, even near the separatrix.  In \citet{Chan2006} we applied our
spectral algorithm to self-gravitating disks and studied Toomre's
criterion. In this section, we use our MHD spectral algorithm to study
the properties of accretion disks driven by the magnetorotational
instability (MRI)~\citep{Balbus1991a}.

Any ionized and magnetized accretion disk is unstable to the MRI as
long as the angular velocity of the flow is a decreasing function of
radius. If a cylindrical disk is threaded by a mean vertical magnetic
field $B_{\rm z}$, all the waves along the vertical direction with
wavenumbers less than \begin{equation} k_\mathrm{MRI} = \sqrt{2q}\
\frac{\Omega}{v_\mathrm{A}}, \label{eq:k_BH}
\end{equation}
are unstable. Here, $\Omega$ is the angular frequency,
\begin{equation}
  q \equiv - \frac{d\ln\Omega}{d\ln r}
\end{equation}
is a measure of the shear, and $v_\mathrm{A} = B_z/\sqrt{4\pi\rho}$ is
the Alfv\'en velocity for the background field.  The growth rate of
the instability depends strongly on the wavenumber and is given
by~\citep{Balbus1991a}
\begin{equation}
  \gamma = \left[- (2-q) - k^2 + \sqrt{(2-q)^2 + 4k^2}\;\right]^{1/2}\Omega.
\end{equation}
The fastest growing mode has a wavenumber given by
\begin{equation}
  k_\mathrm{peak} = \frac{q}{2}\left(\frac{4}{q} - 1\right)^{1/2}
  \frac{\Omega}{v_\mathrm{A}}\;. \label{eq:k_max}
\end{equation}

\begin{figure}
  \plotone{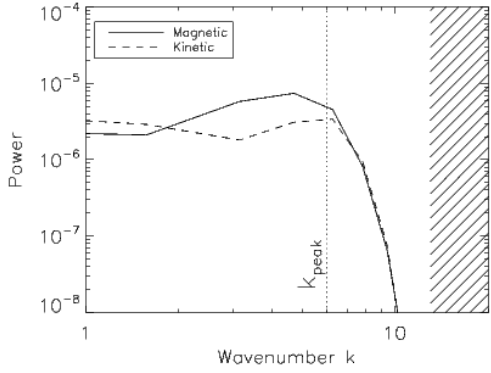}\\
  \caption{The power spectrum along the vertical direction of the
    magnetic and kinetic turbulent energies in a Keplerian disk at
    $r=20$, during the exponential growth of the MRI (at $t=200$), for
    the simulation discussed in the text. The hash-filled area
    represents the limit of the numerical resolution. The vertical
    dotted line corresponds to the wavenumber of the most unstable
    mode as predicted by the analysis of the linear MRI.}
    \label{fig:MRI_ps}
\end{figure}

We use our numerical algorithm to simulate the evolution of a
Keplerian disk in a pseudo-Newtonian potential around a non-rotating
object of mass $M$.  For this calculation, we set $G=c=1$ and all the
distances and times in gravitational units. We solve the MHD equations
in the computational domain $[3, 43] \times [-\pi, \pi) \times [-2,
2)$ using $257 \times 64 \times 32$ grid points.  In order to justify
the periodic boundary conditions along the vertical direction and
avoid numerical problems around the innermost stable circular orbit,
we set the sound speed to $c_\mathrm{s}^2 \approx 0.2$.  We set the
initial density to unity everywhere in the disk and thread the flow
with a vertical magnetic field with a corresponding Alfv\'en velocity
equal to $2\times 10^{-3}$.  In our dimensionless units, it takes a
time of $t \approx 60$ and $t \approx 500$, respectively, for the
fluid elements to complete one orbit at $r = 6$ and $r = 20$.

The initial exponential growth of the MRI offers another possibility
to test the ability of our numerical method to capture the properties
of an MHD instability. For this reason, we discuss first the initial
stages of the simulation and then the properties of the final state of
saturated MHD turbulence.

In Figure~\ref{fig:MRI_ps}, we show the power spectrum of the magnetic
and kinetic energies in the simulation, at a radius $r=20$, during the
initial linear regime of the instability. The similarity of the peak
of the numerical power spectrum to the wavenumber of the most unstable
mode of the linear MRI demonstrates that our numerical algorithm can
reproduce, with an uncertainty comparable to the resolution, the
wavenumber of the mode that shows the fastest growth rate
(eq.~[\ref{eq:k_max}]).

\begin{figure}
  \plotone{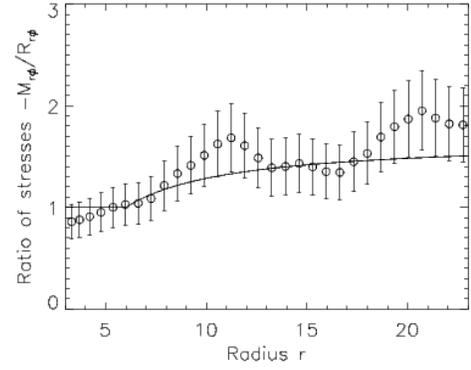}\\
  \caption{The ratio of the $r$-$\phi$ components of the Maxwell, and
    Reynolds stresses, as a function of radius in the accretion flow
    at time $t=2000$.  The error bars indicate the $\simeq 20$\%
    uncertainty in the value of each stress. The solid line shows the
    result of the analytic calculation for the stress
    ratio~\citep{Pessah2006}.}
    \label{fig:MRI_ratio}
\end{figure}

\begin{figure*}
  \plotone{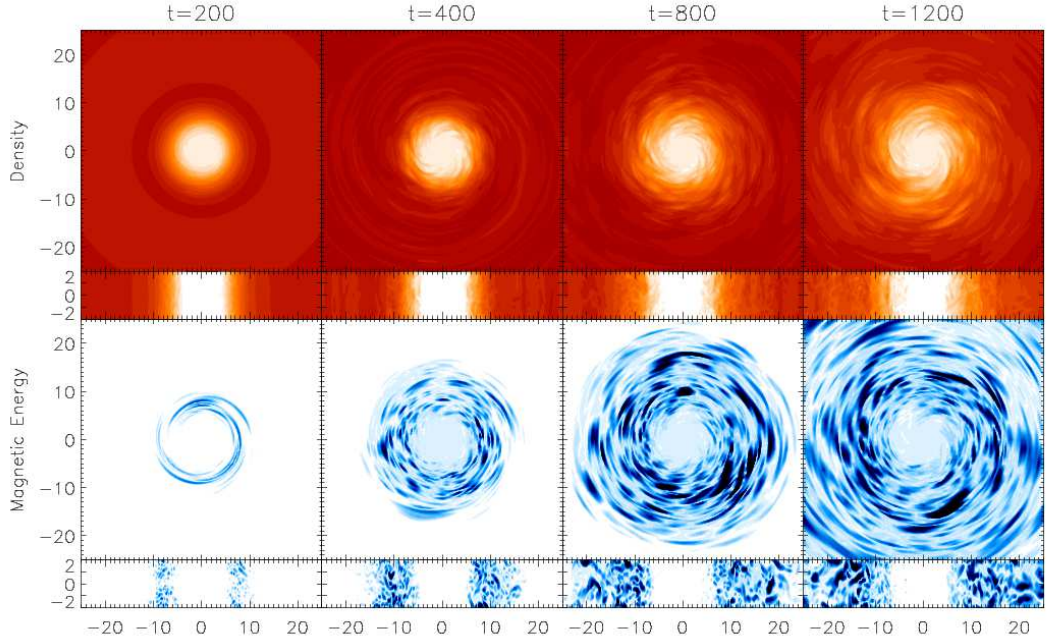}\\ 
  \caption{Four density (upper rows) and magnetic energy (lower rows)
    snapshots of the evolution of the MHD turbulence in an accretion
    disk driven by the magnetorotational instability. The face-on
    views show the two physical quantities at the midplane of the
    domain of solution, whereas the edge-on views are for an azimuth
    of $\phi=0$.  In each row, the first panel corresponds to a state
    very close to the initial laminar flow; the second panel shows the
    density during the exponential growth of the MRI; the last two
    panels show two different instants during the saturated state of
    the turbulence.}
  \label{fig:snapshots}
  \mbox{}

  \mbox{}
\end{figure*}

\begin{figure}
  \plotone{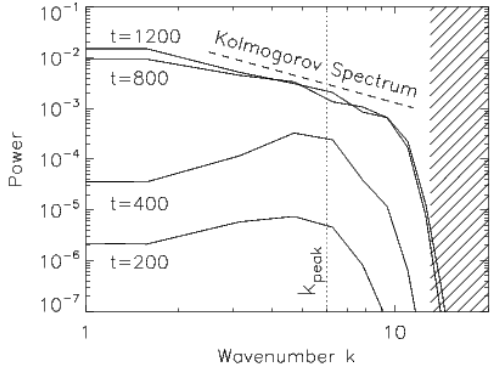}\\
  \caption{The evolution of the power spectrum along the vertical
    direction of the magnetic energy at cylindrical radius $r=20$, for
    the accretion disk simulation shown in Figure~\ref{fig:snapshots}.
    The hash-filled area starts at the cut-off wavenumber of the
    spectral filter and represents the numerical resolution. The
    vertical dotted line corresponds to the wavenumber of the most
    unstable mode as predicted by the analysis of the linear MRI.}
  \label{fig:pds_evol}
  \mbox{}
\end{figure}

The Maxwell and Reynolds stresses are amplified exponentially during
the initial growth of the instability, with their ratios determined
only by the value of local shear~\citep{Pessah2006}.  In
Figure~\ref{fig:MRI_ratio}, we plot the ratio of the Maxwell to
Reynolds stresses as a function of radius in the accretion disk at
time $t=200$ and compare it to the analytical solution
of~\citet{Pessah2006}.  The numerical and analytical ratios agree
well, within the uncertainties of estimating appropriate averages.

After a few orbital periods, turbulence is generated and the solution
of the system of MHD equations becomes highly non-linear.
Figure~\ref{fig:snapshots} shows the evolution of the density and
magnetic energy in the accretion flow from the initial laminar state
(first panel), through the time of the exponential growth of the MRI
(second panel), to the final saturated state of MHD turbulence (last
two panels). As also found in previous simulations of MRI-driven
turbulent accretion disks, the solution is highly variable, with large
fluctuations in the various physical quantities. Moreover, the
magnetic and material stresses are not consistent with the prediction
of the alpha model and are non-vanishing inside the innermost stable
circular orbit. In this paper, we are particularly interested in the
evolution and properties of the power spectrum of turbulence, which is
a statistic that spectral methods are primarily suited to calculate.

Figure~\ref{fig:pds_evol} shows the evolution of the power spectrum
along the vertical direction of the magnetic energy at cylindrical
radius $r=20$.  At early times, the magnetic energy increases
exponentially because of the growth of the MRI, generating a spectrum
of fluctuations that peaks around the wavenumber of maximum growth. At
later times, interactions between different modes lead to spreading of
turbulent energy to nearby modes, generating a power-law spectrum of
fluctuations.  The index of the power spectrum of magnetic energy
fluctuations is comparable to that of a Kolmogorov spectrum, even
though the MHD turbulence is highly anisotropic.

The power-law spectrum of magnetic energy fluctuations extends from
the largest vertical scale of the simulation (which is the vertical
extent of the domain of solution) to the smallest vertical scale
(which is equal to the numerical resolution).  Moreover, the integral
of the power spectrum, which is equal to the total variance, depends
on both scales.  This is consistent with the fact that the saturation
predictor found in numerical simulations of shearing boxes depends on
both the vertical size of the box and the numerical
resolution~\citep{Hawley1995, Hawley1996, Pessah2006b}. It is,
however, a result of two simplifying assumptions in our
simulation. The power spectrum extends to the largest vertical scale
because we have neglected the vertical component of gravity and,
therefore, the disk is not stratified.  At the same time, the power
spectrum extends to the smallest resolved scale because we have
neglected Ohmic dissipation.  Obtaining a realistic saturation
predictor of MRI-driven turbulence in accretion disks will require
numerical simulations of stratified flows with the largest possible
dynamical range and an accurate model of sub-grid physics.


\acknowledgements

C.-K.\,C. and D.\,P.\ acknowledge support from the NASA ATP grant
NAG-513374.


\appendix

\section{Advective-Conservative Mixed Formalism for MHD}
\label{app:scheme}

As in \citet{Chan2005}, we use the non-linear terms $(\mathbf
v\cdot\nabla)\mathbf v$ in their advective forms, whereas we use the
terms that involve the density, $\rho$, and the energy, $E$, in
conservative form.  For the vector potential, we use an advective form
in order to increase stability. In detail:
\begin{eqnarray}
  \partial_t\rho & = & - \frac{\partial_r(r\rho v_r)}{r} -
  \frac{\partial_\phi(\rho v_\phi)}{r} - \partial_z(\rho v_z),
  \label{Eq:hydro_begin}\\
  \partial_t v_r & = & - v_r\partial_r v_r -
  \frac{v_\phi}{r}(\partial_\phi v_r - v_\phi) - v_z\partial_z v_r +
  \frac{\partial_r(r\tau_{rr} - rP - rB^2/8\pi)}{r\rho} +
  \frac{\partial_\phi\tau_{\phi r}}{r\rho} +
  \frac{\partial_z\tau_{zr}}{\rho} + \frac{P + B^2/8\pi -
  \tau_{\phi\phi}}{r\rho} \nonumber\\
  & & + \frac{B_r\partial_r B_r}{4\pi\rho} +
  \frac{B_\phi(\partial_\phi B_r - B_\phi)}{4\pi r\rho} +
  \frac{B_z\partial_z B_r}{4\pi\rho} + g_r \\
  \partial_t v_\phi & = & - v_r\partial_r v_\phi -
  \frac{v_\phi}{r}(\partial_\phi v_\phi + v_r) - v_z\partial_z v_\phi
  + \frac{\partial_r(r\tau_{r\phi})}{r\rho} +
  \frac{\partial_\phi(\tau_{\phi\phi} - P - B^2/8\pi)}{r\rho} +
  \frac{\partial_z\tau_{z\phi}}{\rho}+ \frac{\tau_{r\phi}}{r\rho}
  \nonumber\\
  & & + \frac{B_r\partial_r B_\phi}{4\pi\rho} +
  \frac{B_\phi(\partial_\phi B_\phi + B_r)}{4\pi r\rho} +
  \frac{B_z\partial_z B_\phi}{4\pi\rho} + g_\phi, \\
  \partial_t v_z & = & - v_r\partial_r v_z -
  \frac{v_\phi}{r}\partial_\phi v_z - v_z\partial_z v_z +
  \frac{\partial_r(r\tau_{rz})}{r\rho} + \frac{\partial_\phi\tau_{\phi
  z}}{r\rho} + \frac{\partial_z(\tau_{zz} - P - B^2/8\pi)}{r\rho}
  \nonumber\\
  & & + \frac{B_r\partial_r B_z}{4\pi\rho} + \frac{B_\phi\partial\phi
  B_z}{4\pi r\rho} + \frac{B_z\partial_z B_z}{4\pi\rho} + g_z, \\
  \partial_t E & = & - \frac{\partial_r(rE v_r + rq_r + rF_r)}{r} -
  \frac{\partial_\phi(E v_\phi + q_\phi + F_\phi)}{r} - \partial_z(E
  v_z + q_z + F_z)
  - P\left(\partial_r v_r + \frac{\partial_\phi v_\phi + v_r}{r} +
  \partial_z v_z\right) + \Phi_\nu + \Phi_\eta , \\
  \partial_t A_r & = & v_\phi B_z - v_z B_\phi +
  \eta\left[\frac{1}{r}\partial_r(r\partial_r A_r) +
  \frac{1}{r^2}\partial_\phi^2 A_r + \partial_z^2A_r -
  \frac{A_r}{r^2}\right]
  - \partial_r\dot\Lambda \\
  \partial_t A_\phi & = & v_z B_r - v_r B_z +
  \eta\left[\partial_z^2A_\phi - \frac{A_\phi}{r^2} +
  \frac{1}{r}\partial_r(r\partial_r A_\phi) +
  \frac{1}{r^2}\partial_\phi^2 A_\phi\right]
  - \frac{1}{r}\partial_\phi\dot\Lambda \\
  \partial_t A_z & = & v_r B_\phi - v_\phi B_r +
  \eta\left[\frac{1}{r}\partial_r(r\partial_r A_z) +
  \frac{1}{r^2}\partial_\phi^2 A_z + \partial_z^2A_z\right]
  - \partial_z\dot\Lambda 
  \label{Eq:hydro_end}
\end{eqnarray}
The notation $\partial_t$ denotes partial derivatives with respect to
time. We integrate this equations forward in time using a low storage,
third-order Runge-Kutta scheme \citep[see][for
detail]{Chan2005,Chan2006}.

We compute the magnetic field from the vector potential as
\begin{eqnarray}
  B_r & = & \frac{\partial_\phi A_z}{r} - \partial_z A_\phi \\ B_\phi
  & = & \partial_z A_r - \partial_r A_z \\ B_z & = & \frac{\partial_r
  (rA_\phi)}{r} - \frac{\partial_\phi A_r}{r}
\end{eqnarray}
The only non-trivial term here is $\partial_r(r A_\phi)$, which is in
conservative form.

The viscosity tensor $\tau_{ij}$ in the above equation has the
following general form
\begin{equation}
  \tau_{ij} = 2(\mu_r + \mu_s)e_{ij} + \left(\mu_r + \mu_b -
  \frac{2}{3}\mu_s\right)\left(\nabla\cdot\mathbf v\right)\delta_{ij}.
\end{equation}
As we describe before, $\mu_r$, $\mu_b$, and $\mu_s$ are the
coefficients of radiative, bulk, and shearing viscosity.  The strain
rate tensor $e_{ij}$ written in cylindrical coordinate becomes
\begin{eqnarray}
  e_{rr} & = & \partial_r v_r\\ e_{\phi\phi} & = & \frac{\partial_\phi
  v_\phi}{r} + \frac{v_r}{r} \\ e_{zz} & = & \partial_z v_z\\
  e_{r\phi} = e_{\phi r} & = & \frac{1}{2}\left(\partial_r v_\phi -
  \frac{v_\phi}{r} + \frac{1}{r}\partial_\phi v_r\right) \\ e_{\phi z}
  = e_{z\phi} & = & \frac{1}{2}\left(\frac{\partial_\phi v_z}{r} +
  \partial_z v_\phi\right) \\ e_{zr} = e_{rz} & = &
  \frac{1}{2}(\partial_z v_r + \partial_r v_z)\;.
\end{eqnarray}
The viscous dissipation is
\begin{equation}
  \Phi_\nu = 2(\mu_\mathrm{r} + \mu_\mathrm{s})(e_{ij})^2 +
  \left(\mu_\mathrm{r} + \mu_\mathrm{b} -
  \frac{2}{3}\mu_\mathrm{s}\right)(\nabla\cdot\mathbf v)^2,
\end{equation}
and the Ohmic dissipation rate $\Phi_\eta$ is given by
\begin{equation}
  \Phi_\eta = \frac{\eta}{4\pi}\left[\left(\frac{\partial_\phi B_z}{r}
  - \partial_z B_\phi\right)^2 + \left(\partial_z B_r - \partial_r
  B_z\right)^2 + \left(\frac{\partial_\phi(rB_\phi)}{r} -
  \frac{\partial_\phi B_r}{r}\right)^2\right].
\end{equation}


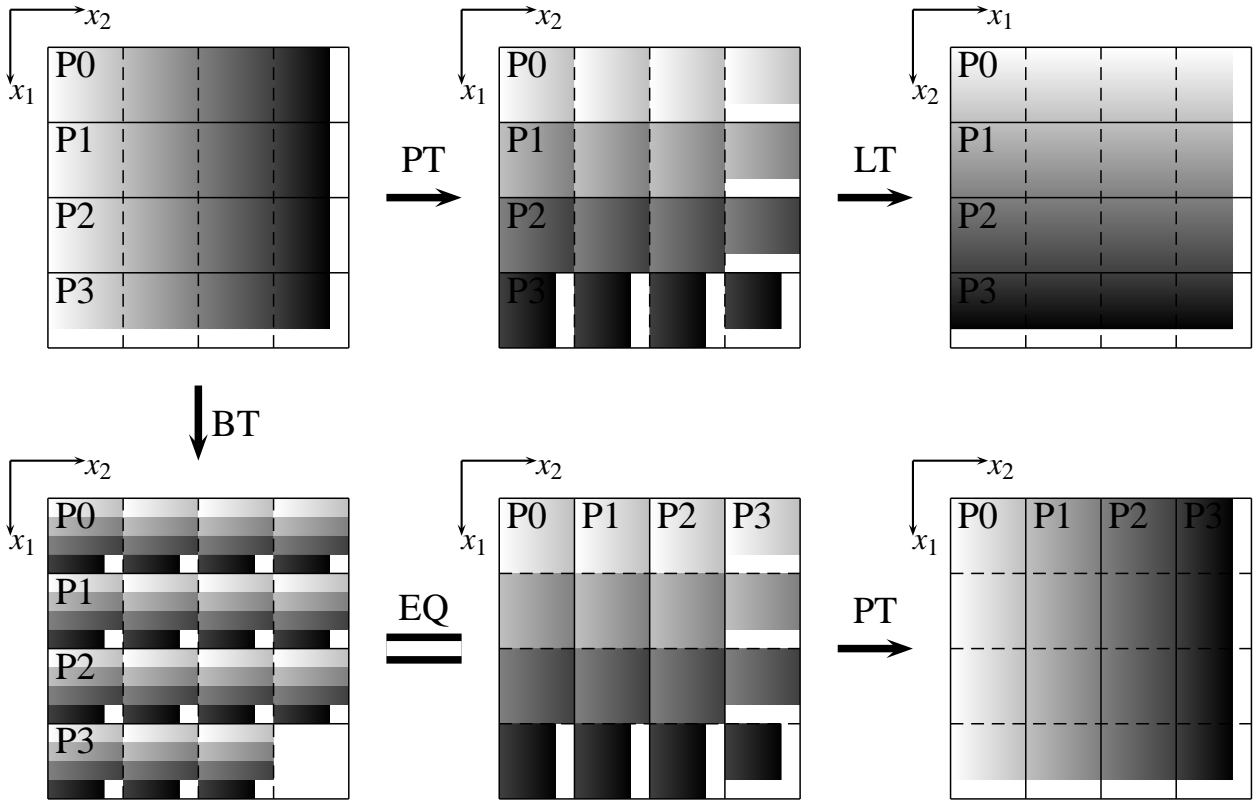
\begin{figure*}
  \begin{pspicture}(18,10.5)
    {\large
      \psline{->}(0.5,10.5)(1.5,10.5)\rput[tl](1.5,10.5){$x_2$}
      \psline{->}(0.5,10.5)(0.5,9.5)\rput[tl](0.5,9.5){$x_1$}
    }
    \psframe[linestyle=none,linewidth=0mm,
      fillstyle=gradient,gradangle=90,gradmidpoint=1,
      gradbegin=white,gradend=lightgray](1,6.25)(2,10)
    \psframe[linestyle=none,linewidth=0mm,
      fillstyle=gradient,gradangle=90,gradmidpoint=1,
      gradbegin=lightgray,gradend=gray](2,6.25)(3,10)
    \psframe[linestyle=none,linewidth=0mm,
      fillstyle=gradient,gradangle=90,gradmidpoint=1,
      gradbegin=gray,gradend=darkgray](3,6.25)(4,10)
    \psframe[linestyle=none,linewidth=0mm,
      fillstyle=gradient,gradangle=90,gradmidpoint=1,
      gradbegin=darkgray,gradend=black](4,6.25)(4.75,10)
    \multido{\i=6+1}{5}{\psline[linewidth=0.2mm](1,\i)(5,\i)}
    \psline[linewidth=0.2mm](1,6)(1,10)
    \multido{\i=2+1}{3}{\psline[linestyle=dashed,linewidth=0.2mm](\i,6)(\i,10)}
    \psline[linewidth=0.2mm](5,6)(5,10)
    {\Large
      \multido{\d=6.6+1,\i=3+-1}{4}{\rput[bl](1.1,\d){P\i}}
    }
    {\large
      \psline{->}(6.5,10.5)(7.5,10.5)\rput[tl](7.5,10.5){$x_2$}
      \psline{->}(6.5,10.5)(6.5,9.5)\rput[tl](6.5,9.5){$x_1$}
    }
    \multido{\il=7+1,\ir=8+1,\d=7.75+1}{3}{
      \psframe[linestyle=none,linewidth=0mm,
	fillstyle=gradient,gradangle=90,gradmidpoint=1,
	gradbegin=white,gradend=lightgray](\il,9)(\ir,10)
      \psframe[linestyle=none,linewidth=0mm,
	fillstyle=gradient,gradangle=90,gradmidpoint=1,
	gradbegin=lightgray,gradend=gray](\il,8)(\ir,9)
      \psframe[linestyle=none,linewidth=0mm,
	fillstyle=gradient,gradangle=90,gradmidpoint=1,
	gradbegin=gray,gradend=darkgray](\il,7)(\ir,8)
      \psframe[linestyle=none,linewidth=0mm,
	fillstyle=gradient,gradangle=90,gradmidpoint=1,
	gradbegin=darkgray,gradend=black](\il,6)(\d,7)
    }
    \psframe[linestyle=none,linewidth=0mm,
      fillstyle=gradient,gradangle=90,gradmidpoint=1,
      gradbegin=white,gradend=lightgray](10,9.25)(11,10)
    \psframe[linestyle=none,linewidth=0mm,
      fillstyle=gradient,gradangle=90,gradmidpoint=1,
      gradbegin=lightgray,gradend=gray](10,8.25)(11,9)
    \psframe[linestyle=none,linewidth=0mm,
      fillstyle=gradient,gradangle=90,gradmidpoint=1,
      gradbegin=gray,gradend=darkgray](10,7.25)(11,8)
    \psframe[linestyle=none,linewidth=0mm,
      fillstyle=gradient,gradangle=90,gradmidpoint=1,
      gradbegin=darkgray,gradend=black](10,6.25)(10.75,7)
    \multido{\i=6+1}{5}{\psline[linewidth=0.2mm](7,\i)(11,\i)}
    \psline[linewidth=0.2mm](7,6)(7,10)
    \multido{\i=8+1}{3}{\psline[linestyle=dashed,linewidth=0.2mm](\i,6)(\i,10)}
    \psline[linewidth=0.2mm](11,6)(11,10)
    {\Large
      \multido{\d=6.6+1,\i=3+-1}{4}{\rput[bl](7.1,\d){P\i}}
    }
    {\large
      \psline{->}(12.5,10.5)(13.5,10.5)\rput[tl](13.5,10.5){$x_1$}
      \psline{->}(12.5,10.5)(12.5,9.5)\rput[tl](12.5,9.5){$x_2$}
    }
    \psframe[linestyle=none,linewidth=0mm,
      fillstyle=gradient,gradangle=0,gradmidpoint=1,
      gradbegin=white,gradend=lightgray](13,9)(16.75,10)
    \psframe[linestyle=none,linewidth=0mm,
      fillstyle=gradient,gradangle=0,gradmidpoint=1,
      gradbegin=lightgray,gradend=gray](13,8)(16.75,9)
    \psframe[linestyle=none,linewidth=0mm,
      fillstyle=gradient,gradangle=0,gradmidpoint=1,
      gradbegin=gray,gradend=darkgray](13,7)(16.75,8)
    \psframe[linestyle=none,linewidth=0mm,
      fillstyle=gradient,gradangle=0,gradmidpoint=1,
      gradbegin=darkgray,gradend=black](13,6.25)(16.75,7)
    \multido{\i=6+1}{5}{\psline[linewidth=0.2mm](13,\i)(17,\i)}
    \psline[linewidth=0.2mm](13,6)(13,10)
   \multido{\i=14+1}{3}{\psline[linestyle=dashed,linewidth=0.2mm](\i,6)(\i,10)}
    \psline[linewidth=0.2mm](17,6)(17,10)
    {\Large
      \multido{\d=6.6+1,\i=3+-1}{4}{\rput[bl](13.1,\d){P\i}}
    }
    {\large
      \psline{->}(0.5,4.5)(1.5,4.5)\rput[tl](1.5,4.5){$x_2$}
      \psline{->}(0.5,4.5)(0.5,3.5)\rput[tl](0.5,3.5){$x_1$}
    }
    \multido{\da=1+1,\db=1.25+1,\dc=1.5+1,\dd=1.75+1,\de=2+1}{3}{
      \multido{\il=1+1,\ir=2+1,\d=1.75+1}{3}{
	\psframe[linestyle=none,linewidth=0mm,
	  fillstyle=gradient,gradangle=90,gradmidpoint=1,
	  gradbegin=white,gradend=lightgray](\il,\dd)(\ir,\de)
	\psframe[linestyle=none,linewidth=0mm,
	  fillstyle=gradient,gradangle=90,gradmidpoint=1,
	  gradbegin=lightgray,gradend=gray](\il,\dc)(\ir,\dd)
	\psframe[linestyle=none,linewidth=0mm,
	  fillstyle=gradient,gradangle=90,gradmidpoint=1,
	  gradbegin=gray,gradend=darkgray](\il,\db)(\ir,\dc)
	\psframe[linestyle=none,linewidth=0mm,
	  fillstyle=gradient,gradangle=90,gradmidpoint=1,
	  gradbegin=darkgray,gradend=black](\il,\da)(\d,\db)
      }
      \psframe[linestyle=none,linewidth=0mm,
	fillstyle=gradient,gradangle=90,gradmidpoint=1,
	gradbegin=white,gradend=lightgray](4,\dd)(5,\de)
      \psframe[linestyle=none,linewidth=0mm,
	fillstyle=gradient,gradangle=90,gradmidpoint=1,
	gradbegin=lightgray,gradend=gray](4,\dc)(5,\dd)
      \psframe[linestyle=none,linewidth=0mm,
	fillstyle=gradient,gradangle=90,gradmidpoint=1,
	gradbegin=gray,gradend=darkgray](4,\db)(5,\dc)
      \psframe[linestyle=none,linewidth=0mm,
	fillstyle=gradient,gradangle=90,gradmidpoint=1,
	gradbegin=darkgray,gradend=black](4,\da)(4.75,\db)
    }
    \multido{\il=1+1,\ir=2+1,\d=1.75+1}{3}{
      \psframe[linestyle=none,linewidth=0mm,
	fillstyle=gradient,gradangle=90,gradmidpoint=1,
	gradbegin=white,gradend=lightgray](\il,0.75)(\ir,1)
      \psframe[linestyle=none,linewidth=0mm,
	fillstyle=gradient,gradangle=90,gradmidpoint=1,
	gradbegin=lightgray,gradend=gray](\il,0.5)(\ir,0.75)
      \psframe[linestyle=none,linewidth=0mm,
	fillstyle=gradient,gradangle=90,gradmidpoint=1,
	gradbegin=gray,gradend=darkgray](\il,0.25)(\ir,0.5)
      \psframe[linestyle=none,linewidth=0mm,
	fillstyle=gradient,gradangle=90,gradmidpoint=1,
	gradbegin=darkgray,gradend=black](\il,0)(\d,0.25)
    }
    \multido{\i=0+1}{5}{\psline[linewidth=0.2mm](1,\i)(5,\i)}
    \psline[linewidth=0.2mm](1,0)(1,4)
    \multido{\i=2+1}{3}{\psline[linestyle=dashed,linewidth=0.2mm](\i,0)(\i,4)}
    \psline[linewidth=0.2mm](5,0)(5,4)
    {\Large
      \multido{\d=0.6+1,\i=3+-1}{4}{\rput[bl](1.1,\d){P\i}}
    }
    {\large
      \psline{->}(6.5,4.5)(7.5,4.5)\rput[tl](7.5,4.5){$x_2$}
      \psline{->}(6.5,4.5)(6.5,3.5)\rput[tl](6.5,3.5){$x_1$}
    }
    \multido{\il=7+1,\ir=8+1,\d=7.75+1}{3}{
      \psframe[linestyle=none,linewidth=0mm,
	fillstyle=gradient,gradangle=90,gradmidpoint=1,
	gradbegin=white,gradend=lightgray](\il,3)(\ir,4)
      \psframe[linestyle=none,linewidth=0mm,
	fillstyle=gradient,gradangle=90,gradmidpoint=1,
	gradbegin=lightgray,gradend=gray](\il,2)(\ir,3)
      \psframe[linestyle=none,linewidth=0mm,
	fillstyle=gradient,gradangle=90,gradmidpoint=1,
	gradbegin=gray,gradend=darkgray](\il,1)(\ir,2)
      \psframe[linestyle=none,linewidth=0mm,
	fillstyle=gradient,gradangle=90,gradmidpoint=1,
	gradbegin=darkgray,gradend=black](\il,0)(\d,1)
    }
    \psframe[linestyle=none,linewidth=0mm,
      fillstyle=gradient,gradangle=90,gradmidpoint=1,
      gradbegin=white,gradend=lightgray](10,3.25)(11,4)
    \psframe[linestyle=none,linewidth=0mm,
      fillstyle=gradient,gradangle=90,gradmidpoint=1,
      gradbegin=lightgray,gradend=gray](10,2.25)(11,3)
    \psframe[linestyle=none,linewidth=0mm,
      fillstyle=gradient,gradangle=90,gradmidpoint=1,
      gradbegin=gray,gradend=darkgray](10,1.25)(11,2)
    \psframe[linestyle=none,linewidth=0mm,
      fillstyle=gradient,gradangle=90,gradmidpoint=1,
      gradbegin=darkgray,gradend=black](10,0.25)(10.75,1)
    \psline[linewidth=0.2mm](7,0)(11,0)
    \multido{\i=1+1}{3}{\psline[linestyle=dashed,linewidth=0.2mm](7,\i)(11,\i)}
    \psline[linewidth=0.2mm](7,4)(11,4)
    \multido{\i=7+1}{5}{\psline[linewidth=0.2mm](\i,0)(\i,4)}
    {\Large
      \multido{\d=7.1+1,\i=0+1}{4}{\rput[bl](\d,3.6){P\i}}
    }
    {\large
      \psline{->}(12.5,4.5)(13.5,4.5)\rput[tl](13.5,4.5){$x_2$}
      \psline{->}(12.5,4.5)(12.5,3.5)\rput[tl](12.5,3.5){$x_1$}
    }
    \psframe[linestyle=none,linewidth=0mm,
      fillstyle=gradient,gradangle=90,gradmidpoint=1,
      gradbegin=white,gradend=lightgray](13,0.25)(14,4)
    \psframe[linestyle=none,linewidth=0mm,
      fillstyle=gradient,gradangle=90,gradmidpoint=1,
      gradbegin=lightgray,gradend=gray](14,0.25)(15,4)
    \psframe[linestyle=none,linewidth=0mm,
      fillstyle=gradient,gradangle=90,gradmidpoint=1,
      gradbegin=gray,gradend=darkgray](15,0.25)(16,4)
    \psframe[linestyle=none,linewidth=0mm,
      fillstyle=gradient,gradangle=90,gradmidpoint=1,
      gradbegin=darkgray,gradend=black](16,0.25)(16.75,4)
    \psline[linewidth=0.2mm](13,0)(17,0)
   \multido{\i=1+1}{3}{\psline[linestyle=dashed,linewidth=0.2mm](13,\i)(17,\i)}
    \psline[linewidth=0.2mm](13,4)(17,4)
    \multido{\i=13+1}{5}{\psline[linewidth=0.2mm](\i,0)(\i,4)}
    {\Large
      \multido{\d=13.1+1,\i=0+1}{4}{\rput[bl](\d,3.6){P\i}}
    }
    {\Large
      \psline[linewidth=1mm]{->}(5.5,8)(6.5,8)
      \rput(6,8.5){PT}
      \psline[linewidth=1mm]{->}(11.5,8)(12.5,8)
      \rput(12,8.5){LT}
      \psline[linewidth=1mm]{->}(3,5.5)(3,4.5)
      \rput(3.5,5){BT}
      \psline[linewidth=1mm,doubleline=true,doublesep=2mm](5.5,2)(6.5,2)
      \rput(6,2.5){EQ}
      \psline[linewidth=1mm]{->}(11.5,2)(12.5,2)
      \rput(12,2.5){PT}
    }
  \end{pspicture}
  \caption{Schematic representation of the standard parallel FFT
    procedure in two dimensions (upper panel) and the alternate
    flip-flop procedure we have implemented in our algorithm (bottom
    panel).}
  \label{fig:PFFT}
\end{figure*}

\section{Code Parallelization}
\label{sec:mpi}

If the magnetohydrodynamic equations were linear, the parallelization
of the spectral algorithm would be trivial, because the coefficient of
each of the basis polynomials would evolve independently from the
others.  However, the non-linear terms in the equations necessitate
cross-processor communication. As discussed in \citet{Chan2005}, we
incorporate the non-linear terms in the time-stepping by transforming
the relevant physical quantities to coordinate space and evaluating
the non-linear terms there, before transforming them back to spectral
space. The efficient parallelization of our algorithm, therefore,
relies on efficiently implementing a parallel version of the Fast
Fourier Transform (FFT) algorithm for a multi-dimensional set of grid
points. We use the Message Passing Interface (MPI) standard to
parallelize our algorithm \citep[see][]{Gropp1999a,Gropp1999b}.

In Figure~\ref{fig:PFFT} we present schematically an example of our
implementation of the FFT algorithm for transforming the values of a
physical quantity on a two-dimensional grid along the $x_1$ and $x_2$
directions.  The standard algorithm for parallel Fast Fourier
Transform starts with an one-dimensional (``slab'') decomposition to
distribute the data across different processors. We use solid lines in
the figure to denote the distribution of grid points on different
processors, which we label by P0, P1, P2, etc.  It is clear from the
upper-left panel of the figure that applying spectral methods along
the $x_2$ direction is straight forward.  We can simply use the
ordinary FFT for each row locally. However, for performing an FFT
along the $x_1$ direction, communication between the various
processors is necessary. The standard method requires taking first a
parallel transpose (PT), exchanging data between processors, and then
a local transpose (LT), per processor, as illustrated in the upper
panels of Figure~\ref{fig:PFFT}. These two processes exchange the
directions of $x_1$ and $x_2$ and FFTs are then applied on each row
separately.

We developed an alternate (``flip-flop'') procedure to increase the
performance of parallel FFTs in multi-dimensions, by avoiding the time
consuming step of obtaining the local transpose on each processor.
The lower row of Figure~\ref{fig:PFFT} illustrates the flipping
procedure, which involves a block transpose (BT) and a parallel
transpose (PT) operation.  The flopping procedure follows the
opposite routine. In these operations we can take advantage of the
resulting memory layout and use the cache more efficiently. Because
the order of the indices is preserved, the algorithm is easier to
implement. More importantly, depending on the domain size and the
number of processors we use, this approach can speed up the parallel
FFT by $\sim$20\% in our 32-processor Beowulf cluster.


\end{document}